\documentclass[aps,pra,reprint,groupedaddress,showpacs,amsmath,amssymb]{revtex4-1}

\usepackage{graphicx}
\usepackage{dcolumn}
\usepackage{bm}
\usepackage{natbib}

\begin{document}


\title{Phase Coherence and Fragmentation of Two-Component Bose-Einstein Condensates Loaded in State-Dependent Optical Lattices}



\author{Hyunoo Shim}
\affiliation{Department of Physics and Astronomy, Stony Brook University, Stony Brook, New York 11794-3800, USA}
\affiliation{Department of Actuarial Science, Hanyang University, Ansan 15588, Korea}
\author{Thomas Bergeman}%
\affiliation{Department of Physics and Astronomy, Stony Brook University, Stony Brook, New York 11794-3800, USA}


\date{\today}

\begin{abstract}

A binary mixture of interacting Bose-Einstein condensates (BEC) in the presence of
fragmentation-driving external lattice potentials forms
two interdependent mean-field lattices made of each component. These effective mean-field 
lattices, like ordinary optical lattices, can induce additional fragmentation and phase 
coherence loss of BECs between lattice sites. In this study, we consider the nonequilibrium 
dynamics of two hyperfine states of one-dimensional Bose-Einstein condensates, subjected to state-dependent optical lattices. Our numerical calculations using the truncated Wigner approximation (TWA) show that phase coherence in a mixture of two-component BECs can be lost not just by optical lattices, but by mean-field lattices gradually formed by other components, and we reveal that such an effect of internal mean-field lattices, however, is limited, contrary to external optical lattices, in regard to phase decoherence.

\end{abstract}


\maketitle

\section{\label{sec:level1}Introduction}

The properties of ultracold atoms in optical lattices have been studied
intensively as models for various condensed matter
phenomena \cite{jaksch05,lewenstein07}. In many
cases, the coherence or lack of coherence between atoms in adjacent wells
plays a crucial role, especially in the superfluid-Mott insulator phase
transition \cite{fisher89, Jaksch98, greiner02, stoferle04}. For example, the
work of Orzel {\it et al.} \cite{Orzel01}, with a 1D array of ``pancake''
condensates, displayed high-visibility interference patterns under conditions
of phase coherence between adjacent wells, but dramatic reduction of the
interference contrast, or visibility, when the wells were deepened. Later
work with 3D optical lattices by Greiner {\it et al.} \cite{greiner02} 
exhibited the superfluid to Mott-Insulator phase transition through the 
interference pattern when the 3D condensates were released.

More recently, diverse aspects along the phase transition are subject to 
study such as phase diagrams \cite{Polak10}, strong 
interaction \cite{Buonsante08}, and special geometries \cite{Polak08}. Among 
them are experimental studies of systems in which the Bose
condensate consists of ``distinct components'', such as different atomic
species, different substates or different hyperfine levels. These studies 
focus on multi-component 
systems \cite{thalhammer08,catani09,pertot10,hamner11,soltan-panahi11} and 
it is easy to imagine further experimental studies probing the rich physics 
of multi-component BECs. For example, experiments have addressed the question 
of phase coherence of two-component Bose-Einstein condensates, and it has been
observed \cite{catani08} that the presence of $^{41}$K atoms reduces the 
visibility of the interference pattern of marginally-overlapped $^{87}$Rb 
atoms in a 3D optical lattice. Similarly, in a condensate of miscible 
$^{87}$Rb atoms in a state-dependent 3D optical lattice, the presence of 
atoms in a second hyperfine level can reduce the superfluid coherence of 
atoms in a first hyperfine level \cite{gadway10}. Lasers in these studies
were tuned such that both components experience peak-matched lattice potentials.  

In a theoretical perspective, there have been a number of studies addressing
aspects of Bose condensates with such multi-components. Perhaps most notable
have been discussions of different phase regimes and phase 
transitions \cite{altman03,isacsson05,krutitsky03} and of the extended 
Bloch band structure \cite{larson08}. There have also been studies
of dynamical effects including those associated with ramping up the optical
lattice \cite{ruostekoski07,wernsdorfer10,Cipolatti16}. Recent studies show 
theories about, for example, phase diagram and 
stability \cite{Guglielmino10, Li13}, evolution of coherence or 
number-squeezing during ramp-up \cite{Hofer12, Shrestha12, Xi14}. 
Other works have investigated other equilibrium or nonequilibrium properties via various stochastic theories including a truncated Wigner method applied to single-component BECs \cite{Cockburn11, isella06}, which will also be interesting once extended to multi-component BECs.

In many cases, theoretical studies in deep lattices have used the 
Bose-Hubbard 
model (BHM) \cite{altman03,isacsson05,krutitsky03,wernsdorfer10,Cipolatti16,bruderer07} 
or Time-Evolving Bloch Decimation (TEBD) approach \cite{hu09}, both of which 
become problematic when there are many atoms per well, as in the 
one-component experiments of \cite{Orzel01}. Instead, some studies have 
developed novel analytic methods \cite{Zaleski11, Zheng13, Sajna15, Yanay16}. 
On the other hand, for BECs in shallow optical lattices, theoretical analysis 
of phase decoherence in 1D has been extensively performed via an extended 
Gross-Pitaevskii equation (GPE) approach \cite{mckagan06}, or via the 
Truncated Wigner Approximation (TWA) approach \cite{isella06}. Whereas the 
application of the GPE to such systems is limited to shallow lattices and 
low temperatures unless used in a full 3D treatment \cite{mckagan06}, the 
TWA, which evolved from quantum optics applications \cite{gardiner92,walls08} 
has emerged as the promising method for simulation of Bose-Einstein 
condensates in optical lattices.  The TWA has also been used to model 
dephasing of single component BECs in 1D optical 
lattices, \cite{isella06,bistritzer07}. 

In this paper, we study phase decoherence of interpenetrating peak-mismatched 
two-component mixtures that are slowly loaded into relatively shallow 
state-dependent lattices, $V_{o,A}(z,t)=s_A(t) E_R \cos^2(kz)$ for 
component (A) and $V_{o,B}=0$ (or $V_{o,B}(z,t)=s_B(t) E_R \sin^2(kz)$) for 
component (B) ($s_i$ is a scale of lattice height for the component $i$ and 
$E_R$ is a recoil energy) using the TWA. We construct a TWA model for 
two-component BEC clouds which are independently phase-coherent in the 
initial state. We focus on the effects of both components when there is a 
single optical lattice acting on component 
A ($V_{o,A} \propto \cos^2(kz), V_{o,B}=0$) or alternatively when there are 
two half-period mismatched 
optical lattices ($V_{o,A} \propto \cos^2(kz), V_{o,B} \propto \sin^2(kz)$). 
This work is restricted to phenomena at zero temperature
($T$=0) and one dimension (1D).

We find, as in the experimental studies with 3D 
condensates \cite{catani08,gadway10}, that, in the former case ($V_{o,B}=0$), 
the second component diminishes the phase coherence of the first component, 
and also experiences decoherence itself relative to the initial fully 
coherent state, due to formation of atomic mean-field lattices. 
We also find that, in the latter case ($V_{o,B} \propto \sin^2(kz)$), the
effect of an atomic mean-field lattice is limited in reducing phase coherence 
of the other component.

For a qualitative explanation of the fragmentation processes described above,
we adopt a simple Gaussian variational ansatz for single-particle Wannier 
functions.  We find that the model shows a good agreement with the trend of 
fragmentation inferred from the above TWA calculations \cite{isella06}. 

In view of the numerous theoretical and experimental papers on cold atoms
in optical lattices, we stress again that our work extends to two components
the results of \cite{isella06} on quantum fluctuations and phase decoherence.
Also we display explicitly the site to site decoherence due to lattice
ramp changes, summarized in general in \cite{wernsdorfer10}.

The layout of this paper is as follows. In Sec. \ref{sec:dyn_two_BECs}, we 
construct the TWA model for 1D two-component BECs beginning from a 
second-quantized effective Hamiltonian. In Sec. \ref{sec:init_states}, the 
TWA representation is applied to initial states, where the Wigner probability 
distribution for the initial state is found \cite{gardiner04}, and we prepare 
an ensemble of initial states under the Wigner distribution. 
We present the main results of the paper in 
Sec. \ref{sec:results}. Sec. \ref{sec:phcoh_frac}, introduces a single 
state-selective optical lattice and shows the effect of an added component 
on phase coherence loss over a range of populational fractions of each 
component. We also implement a variational ansatz calculation to explain the 
patterns found above. Sec. \ref{both_frag} then continues the similar setup 
but with variable lattice heights to see the fragmentation induced by lattice 
height increase. In Sec. \ref{sec:nonmono}, we find limited fragmentation 
(non-monotonic dependence on lattice heights) as two state-dependent optical 
lattices are turned on.  Finally, Sec. \ref{sec:conclusion} is devoted
to concluding remarks.

\section{Dynamics of two-component BECs in the TWA}\label{sec:dyn_two_BECs}

We consider a mixture of two Bose-Einstein condensates which is
confined in a harmonic trap, where the two components are two hyperfine
states of the same species \cite{hall98}. The harmonic trap
potential is $V_{har}(\vec{x})=\frac{1}{2}m(\omega_z^2z^2+\omega_{\rho}^2
(x^2+y^2))$ with a weak longitudinal trap frequency ($\omega_z$) and a stronger
transverse trap frequency ($\omega_{\rho}$) ($\omega_z < \omega_{\rho}$), so
that the BECs are cigar-shaped.

Assuming effective 1D BECs with negligible transverse excitations, as 
explained in Appendix \ref{sec:num_meth}, the effective 1D two-component 
second-quantized Hamiltonian for the system is
\begin{eqnarray}
  H &=& \sum_{i=A,B}\int dz \hat{\psi}_i^{\dagger}(z)L_{i}\hat{\psi}_i(z)
\nonumber\\
 & & +\frac{1}{2}\sum_{i=A,B}g_{ii}\int dz \hat{\psi}_i^{\dagger}(z)
\hat{\psi}_i^{\dagger}(z)\hat{\psi}_i(z)\hat{\psi}_i(z) \nonumber\\
 & & +g_{AB}\int dz \hat{\psi}_A^{\dagger}(z)\hat{\psi}_B^{\dagger}(z)
\hat{\psi}_A(z)\hat{\psi}_B(z), \label{eq:H_sys}
\end{eqnarray}
and the $L_i$ is defined as,
\begin{eqnarray}
L_i=-\frac{\hbar^2\nabla^2}{2m_{i}}+V_{h,i}(z)+V_{o,i}(z,t)-\mu_i.
\end{eqnarray}

Here, we label the first species as `A' and the second one as `B'. For each
species, $m_i$ is the particle mass, $\mu_i$ is the chemical potential,
$V_{h,i}(z)=m_i\omega_z^2 z^2/2$ is the external harmonic trap potential,
$V_{o,i}(z,t)$ is the time-varying state-dependent optical lattice potential
along the axial direction. For an effective 1D BEC with a Gaussian profile 
along the transverse direction, $g_{ij} = 2\hbar\omega_{\rho}a_{ij}$, 
where $a_{ij}$ is the scattering length, if the two masses are equal.

The equation of motion for the component $i$ field, $\hat{\psi}_i(z)$, is
\begin{eqnarray}
i\frac{d}{dt}\hat{\psi}_i&=&
\hat{H}_i\hat{\psi}_i \equiv L_i\hat{\psi}_i+
\sum_{j}g_{ij}\hat{\psi}_j^{\dagger}\hat{\psi}_j\hat{\psi}_i.
\end{eqnarray}

In Appendix \ref{sec:TWA}, we construct a TWA method for the above 
two-component fields. Then we obtain the corresponding Fokker-Planck equation. 
We can translate such a Fokker-Planck equation into the stochastic 
differential equation for the classical Wigner fields, 
$\psi_{i}(z,t)$ \cite{gardiner04}. The resulting equation for a single 
realization of the Wigner fields that describes a single trajectory in phase 
space is
\begin{eqnarray} \label{dpsidt}
i\hbar \frac{\partial \psi_{i}(z,t)}{\partial t}&=&
[L_i+\sum_{j}g_{ij}(|\psi_{j}(z,t)|^2-d_{ij})] \nonumber\\
& & \psi_{i}(z,t), 
\end{eqnarray}
where $d_{ij} = 1 \phantom{a} (\mathrm{or} \phantom{a} 1/2)$ if 
$i=j \phantom{a} (\mathrm{or} \phantom{a} i \ne j)$. 

Since the third-order diffusion process is neglected, the stochastic 
fluctuations during the time evolution are absent, but the initial state 
still has quantum fluctuations following the probability distribution given 
by the Wigner representation \cite{gardiner04}. Therefore, given the initial 
condition for each realization following the Wigner function, the classical 
field, $\psi_i(z,t)$, evolves under the above deterministic trajectory which 
resembles the Gross-Pitaevskii equation except the small depletion terms 
indicated by the ``$-d_{ij}$'' quantities in Eq. (\ref{dpsidt}).

\section{Stochastic initial states and phase coherence between 
sites}\label{sec:init_states}

The system we discuss is a two-component 1D BEC confined by the same
harmonic trap. In this discussion, we consider the two hyperfine
states of $^{87}$Rb atoms, $\left|F=1,m_F=-1 \right\rangle$ and 
$\left|F=2,m_F=-2 \right\rangle$. Since the differences between interaction 
strengths ($a_{AA},a_{BB},a_{AB}$) are small,
we assume that the two atoms share the same intraspecies and interspecies
interaction strength ($a_s \equiv a_{AA}=a_{BB}=a_{AB}=5.5nm$) and they
have the same masses ($m=m_{A}=m_{B}$).

For BECs with a
large number of atoms at sufficiently low temperatures ($T \ll T_c$), the
Bogoliubov quasiparticle description \cite{ozeri05} is a good approximation
to the exact many-body dynamics of the system, provided that the number of
noncondensate particles ($N_{ex}$) is sufficiently smaller than that of 
condensate atoms ($N_c$) ($N_{ex} \ll N_c$) \cite{blakie08}. A more exact 
number-conserving theory would be based on an expansion in powers of
$1/\sqrt{N}$ \cite{sinatra02,sinatra01,gardiner97} using the
Particle-Number Conserving formalism (PNC) \cite{castin97a,castin98}.

In the Bogoliubov theory, the matter-wave field operator, in addition to the
condensate field operator, includes small quasiparticle amplitudes,
\begin{eqnarray}
\hat{\psi}_i(z)&=&\psi_{i0}(z)\hat{\alpha}_{i0} + \sum_{\mu > 0}(u_{i\mu}(z)
\hat{\alpha}_{\mu}-v_{i\mu}(z)\hat{\alpha}_{\mu}^{\dagger}), 
\end{eqnarray}
where $\hat{\alpha}_{i0}$ is the annihilation operator for the component $i$
condensate mode, whereas $\hat{\alpha}_{\mu}$ is the quasiparticle annihilation
operator for the collective mode $\mu$. These operators satisfy the bosonic
commutation relation, $[\hat{\alpha}_{\mu},\hat{\alpha}_{\nu}^{\dagger}]
=\delta_{\mu\nu}$, etc. The normalization conditions for the single-particle
condensate amplitudes and for the Bogoliubov quasiparticle mode amplitudes
are
\begin{eqnarray}
\int dz \psi_{A0}^*(z)\psi_{A0}(z)=\int dz \psi_{B0}^*(z)\psi_{B0}(z)=1, \\
\int dz \big[u_{A\mu}^*(z)u_{A\nu}(z)+u_{B\mu}^*(z)u_{B\nu}(z)\nonumber\\
-v_{A\mu}^*(z)v_{A\nu}(z)-v_{B\mu}^*(z)v_{B\nu}(z)\big]=\delta_{\mu\nu}.
\end{eqnarray}

The expectation values of the number operator correspond to the populations
in the condensate mode for each component and in the collective modes.
\begin{eqnarray}
\langle\hat{\alpha}_{i0}^{\dagger}\hat{\alpha}_{i0}\rangle&=&N_{i0},\\
\langle\hat{\alpha}_{\mu}^{\dagger}\hat{\alpha}_{\mu}\rangle&=&
n_{\mu}=\frac{1}{\exp{(\epsilon_{\mu}/k_{B}T)}-1},
\end{eqnarray}
where the Bogoliubov quasiparticles are in thermal equilibrium at $T$.

The quasiparticle mode amplitudes, $u_{i\mu},v_{i\mu}$ satisfy the coupled 
Bogoliubov-de Gennes equation for two-component BEC:
\begin{widetext}
\begin{eqnarray}
\left(\begin{array}{cccc}
H_A+h_{AA}|\psi_A|^2 & h_{AB}\psi_A\psi_B^* &
-h_{AA}\psi_A^2 & -h_{AB}\psi_A\psi_B \\
h_{AB}\psi_A^*\psi_B & H_B+h_{BB}|\psi_B|^2 &
-h_{AB}\psi_A\psi_B & -h_{BB}\psi_B^2 \\
-h_{AA}(\psi_A^*)^2 & -h_{AB}\psi_A^*\psi_B^* &
H_A+h_{AA}|\psi_A|^2 & h_{AB}\psi_A^*\psi_B \\
-h_{AB}\psi_A^*\psi_B^* & -h_{BB}(\psi_B^*)^2 &
h_{AB}\psi_A\psi_B^* & H_B+h_{BB}|\psi_B|^2 \\
\end{array}\right)
\left(\begin{array}{c}
u_{A\mu}\\
u_{B\mu}\\
v_{A\mu}\\
v_{B\mu}\\
\end{array}\right)
=
\left(\begin{array}{c}
\epsilon_{\mu} u_{A\mu}\\
\epsilon_{\mu} u_{B\mu}\\
-\epsilon_{\mu} v_{A\mu}\\
-\epsilon_{\mu} v_{B\mu}\\
\end{array}\right), \nonumber\\
\label{eq:BdG}
\end{eqnarray}
\end{widetext}
where $h_{ij}=g_{ij}\sqrt{N_iN_j}$. 

In Appendix \ref{sec:TWA}, we generate classical stochastic fields for the
initial state in the Wigner representation. 
Having prepared such initial stochastic fields and their time evolution, 
we are especially interested in the short-range non-local
coherence of subcondensates between neighboring sites at each time. We define 
a subcondensate projection operator for each site $l$ as in \cite{steel98}:
\begin{eqnarray}
\hat{a}_{il}(t)=\int_{l^{th} 
\mathrm{site}}dz\bar{\psi}_{GP}(z,t)\hat{\psi}_{i}(z,t),
\label{eq:proj_op}
\end{eqnarray}
where $\hat{a}_{il}$ is the annihilation operator for component $i$ in 
the $l^{th}$ well, $\bar{\psi}_{GP}$ the solution of the GPE, normalized to 
one within each well.  The site positions are different for the two 
components as explained below. This operator is defined as a stochastic field 
operator whose amplitudes are projected over the ground state of each 
condensate mode.  The projection method allows us to avoid complicated 
calculations of symmetrically-ordered multimode fields \cite{blakie08}.

In this study, a state-dependent optical lattice for the component $i$ is a
sinusoidal function,
$V_{o,A}(z,t)=s_A(t) E_R \cos^2(kz)$ (and $V_{o,B}(z,t)=s_B(t) E_R 
\sin^2(kz)$ if exists), where $s_i$ is the scale of lattice height for the 
component $i$, and $E_R = \hbar^2k^2/2m$ is the recoil energy with $m=m_A=m_B$.
$\psi_A$ and $\psi_B$ are localized at the odd sites $z=\pm (2n+1)d/2$, at 
the even sites $z=\pm 2nd/2$ ($n=0,1,2,...$), respectively.  Repulsive 
interspecies interactions repel component B atoms from the localization sites
of the component A. 

We now consider moments of the Wigner function of interest. First, the
occupation number of component $i$ in the $l^{th}$ site is
\begin{eqnarray}
n_{il} = \langle\hat{a}_{il}^\dagger\hat{a}_{il}\rangle =
\langle\hat{a}_{il}^\dagger\hat{a}_{il}\rangle_W - \frac{1}{2},
\label{eq:num_exp}
\end{eqnarray}
where $\langle \cdots \rangle_W$ means an expected value in the Wigner 
representation.

The equal-time first-order coherence is the phase coherence of component 
$i$ between two sites at the time $t$:
\begin{eqnarray}
g_i^{(1)} \equiv C_{i;ll'}(t) &=& \frac{|\langle\hat{a}_{il}^\dagger(t)
\hat{a}_{il'}(t)\rangle|}{(\langle\hat{a}_{il}^\dagger(t)\hat{a}_{il}(t)\rangle
\langle\hat{a}_{il'}^\dagger(t)\hat{a}_{il'}(t)\rangle)^{1/2}} \hspace*{5mm} 
\nonumber\\
 &=&  \frac{|\langle\hat{a}_{il}^\dagger\hat{a}_{il'}\rangle|}
{\sqrt{n_{il}n_{il'}}}, 
\end{eqnarray}
where in the last equation the notation is simplified via Eq.
\ref{eq:num_exp}. For brevity, we now omit the time dependence from
the expectation values of the condensate mode operators.

\section{Phase decoherence and fragmentation of two-component
BECs}\label{sec:results}
\subsection{A single lattice (A) with varying fractions of a mixture}\label{sec:phcoh_frac}

We now examine the phase decoherence patterns of two-component BECs at
$T=0$, driven by a single state-dependent optical lattice. 
Both components are trapped by the same anisotropic harmonic potential.  
In this subsection, component A is placed in an optical lattice whereas no 
external lattice is applied to component B.  We fix the total atom number 
and the ramp-up time, and vary the number ratio of A to B atoms. 

Having prepared the initial state of superfluid BECs placed in the
harmonic trap, we linearly turn on the optical lattice up to a final
height of $s_{max,A} = 10$ in ramp-up time of 
$\omega_R\tau_{RU}=250$ ($\omega_R$ is a recoil frequency as defined in 
Appendix \ref{sec:num_meth}),
then maintain the height until the end of simulations:
\begin{eqnarray}
V_{o,A}(z,t) &=& s_A(t) E_R \cos^2(kz), \nonumber\\
V_{o,B}(z,t) &=& 0,
\end{eqnarray}
where $s_A(t) = s_{max,A} t/\tau_{RU}$, $0 \le t \le \tau_{RU}$.
We fix the total number of atoms,  $N_{tot} = 5 \times 10^3$, and vary the 
fractions of components A and B, $f_A$ and $f_B=1-f_A$, in order to see the 
effects of interspecies interaction and imbalanced populations on phase 
decoherence. 
 
First, we remind ourselves of phase decoherence of single-component BECs in 
optical lattices. From previous experimental and theoretical 
work \cite{Orzel01,isella06}, we expect component A in the absence of B atoms 
to exhibit phase decoherence under certain conditions. As the periodic lattice 
rises into the BEC cloud, the regions occupied by the lattice peaks are 
locally avoided by ground state component A and 
the wavefunctions are eventually fragmented to some degree. 
The tunneling rate of the wavefunctions between the 
adjacent sites is reduced so that the fluctuation in each subcondensate 
breaks the long-range phase coherence.  

\begin{figure}
\includegraphics[scale=0.3]{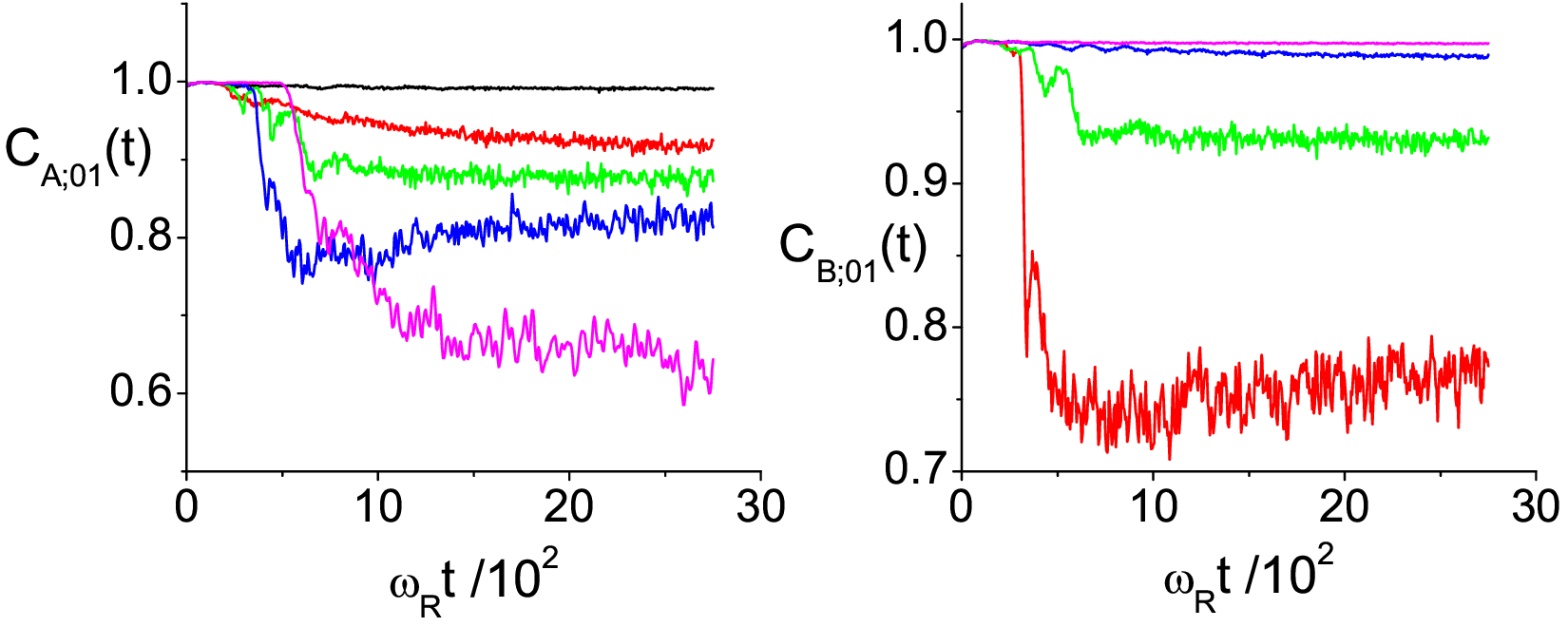}
\caption{(Color online) Phase coherence of component A (on left) and 
component B (on right) for various populational fractions ($0 \le f_B < 1$), 
for $N=5 \times 10^{3}$, $\tau_{RU}=250/\omega_{R}$ = 11ms for $^{87}$Rb. The
numerical simulation shows the results from when lattice loading begins.
The fractions of component B, $f_{B}$, and the line colors are (in the large 
$t$ limit, top to bottom on the left; bottom to top on the right): 0, black 
(for left side only); 0.1, red; 0.2, green; 0.4, blue; 0.6, pink. 
\label{ppl_a1}}
\end{figure}

Figure \ref{ppl_a1} shows the change in phase coherence between the
center and nearest neighbor well, $C_{i;01}$, for component A (left) and 
B (right) from the time the lattice begins to ramp up, to a large time limit.  
The coherence changes
for other distant wells ($C_{i;02}, C_{i;03}, C_{i;04}$, etc.) 
exhibit a similar pattern, but with more coherence loss at a given time.
The first-order correlation functions between sites are closely related
to the visibilities of the interference pattern \cite{ramanan09,greiner02}. 
For one-component BEC cases, a 
complete loss of phase coherence would imply a transition to the Mott 
insulator state. In these calculations, the maximum lattice height does not
reach the Mott insulator regime, as indicated by the observation that
in Fig. \ref{ppl_a1}, $C_{A;01}$ remains very close to unity if $f_{A}=1$.
However, as $f_{B}$ increases, component A exhibits decoherence.

\begin{figure}
\includegraphics[scale=0.30]{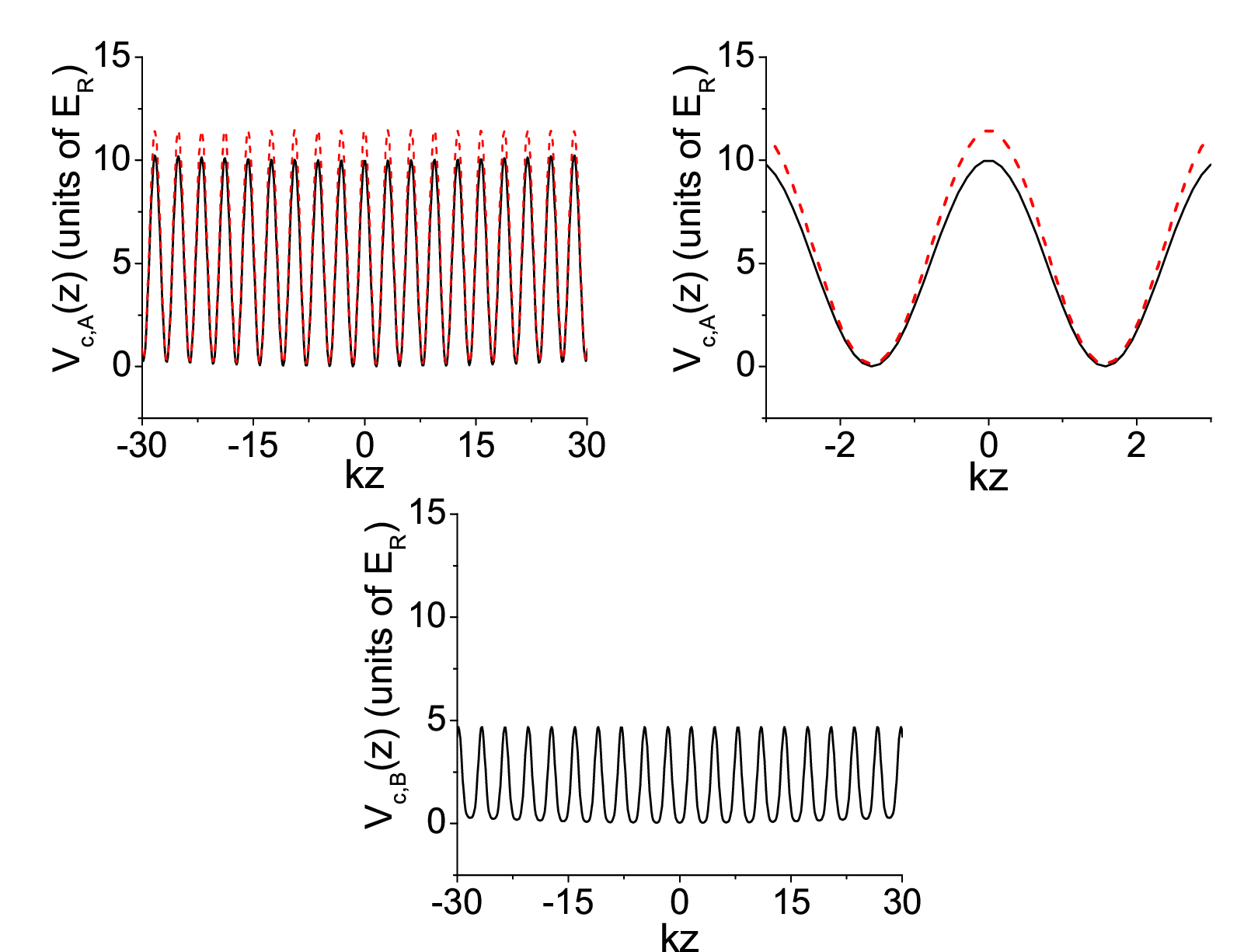}
\caption{(Color online) The profile of combined potentials for the component A,
$V_{c,A}(z)=V_{h,A}(z)+V_{o,A}(z)$ (black solid line on top left) and 
$V_{c,A}(z)=V_{h,A}(z)+V_{o,A}(z)+g_{AB}|\psi_{B}(z)|^2$ (red dashed line on 
top left) with $s_{max,A}=10$, and for the
component B, $V_{c,B}(z)=V_{h,B}(z)+g_{AB}|\psi_{A}(z)|^2$ (on bottom). The 
figure on top right is the same plot on top left, but enlarged near the center 
of trap. The potentials, $V_{c,i}(z)$, are in units of $E_R$. The populations
are $N_A=4 \times 10^3$, $N_B=1 \times 10^3$.\label{ppl_al}}
\end{figure}

A new feature in this two-component case is the reduction of phase coherence 
of component B, which is induced as for component A but with the role of
the optical lattice replaced by the atomic mean-field potential formed by
component A's periodic localization. In the GPE for the component B,
\begin{eqnarray}
i\hbar \frac{\partial \psi_{B}(z,t)}{\partial t}&=& \big[L_B
+g_{BB}|\psi_{B}(z,t)|^2\nonumber\\
& & + g_{AB}|\psi_{A}(z,t)|^2 \big]\psi_{B}(z,t),
\end{eqnarray}
such spatial variation of potential is expressed by the term $g_{AB}|\psi_{A}(x,t)|^2$.
In Fig. \ref{ppl_al}, we show the optical lattice with the harmonic trap,
which directly affects coherence properties of component A, and the
mean-field lattice of A with the same harmonic trap, acting on component B, for
the case $N=5 \times 10^{3}$, $N_{A} = 4 \times 10^{3}$.
The distortion by the harmonic trap potential is almost negligible around
the center. We denote the atomic mean-field lattice made by component A as
\begin{eqnarray}
I_{al,A}(z) = g_{AB} |\psi_{A}(z)|^2.
\end{eqnarray}
Then, the depth of the optical lattice and the interaction strength of
the mean-field lattice are comparable ($(I_{al,A}(z)|_{max}-I_{al,A}(z)
|_{min})/(V_{o,A}(z,t)|_{max}-V_{o,A}(z,t)|_{min}) \simeq 0.6$) for
$f_B/f_{A} = 1/4$ as can be seen by Fig. \ref{ppl_al}.

Due to the presence of the mean-field lattice, the tunneling amplitude between
the localization sites for component B is reduced, resulting in coherence loss,
as shown on the right of Fig. \ref{ppl_a1}. 

The phase decoherence of component A is greater in the presence of 
component B than without component B, and increases as $f_{B}$ increases.
Note also that the mean-field
potential from B atoms acting on A atoms is in phase with the optical lattice,
and thus effectively raises the periodic potential that A atoms see, therefore
contributing to the loss of coherence of the A atoms. However, comparing
with the degree of coherence for A atoms alone as a function of lattice height
shown in the next section, elevation of the effective lattice
acting on A atoms does not explain fully the decrease of coherence shown in 
Fig. \ref{ppl_a1} (left side).  Evidently the stochastic nature of the
atom distributions also plays a role.

The experiment in \cite{gadway10} has shown a similar dependency on 
populational fractions but with two peak-matched state-dependent optical 
lattices in order to place two components at the same lattice site. 

To gain another perspective on these processes, we expand the
wavefunctions in an array of Wannier-like orbitals, $w_i(z)$,
\begin{eqnarray}
\hat{\psi}_i(z) = \sum_{l} \hat{a}_{il} w_i(z-R_{il}),
\end{eqnarray}
where the single particle wavefunction, $w_i(z-R_{il})$ is centered at
$R_{Al} = (2l\pm 1)d/2$, $R_{Bl}=2ld/2$ for each component. We can 
approximate the Wannier functions as Gaussian functions and calculate on-site 
interaction energies and widths of on-site single-particle wavefunctions.

\begin{figure}
\includegraphics[scale=0.30]{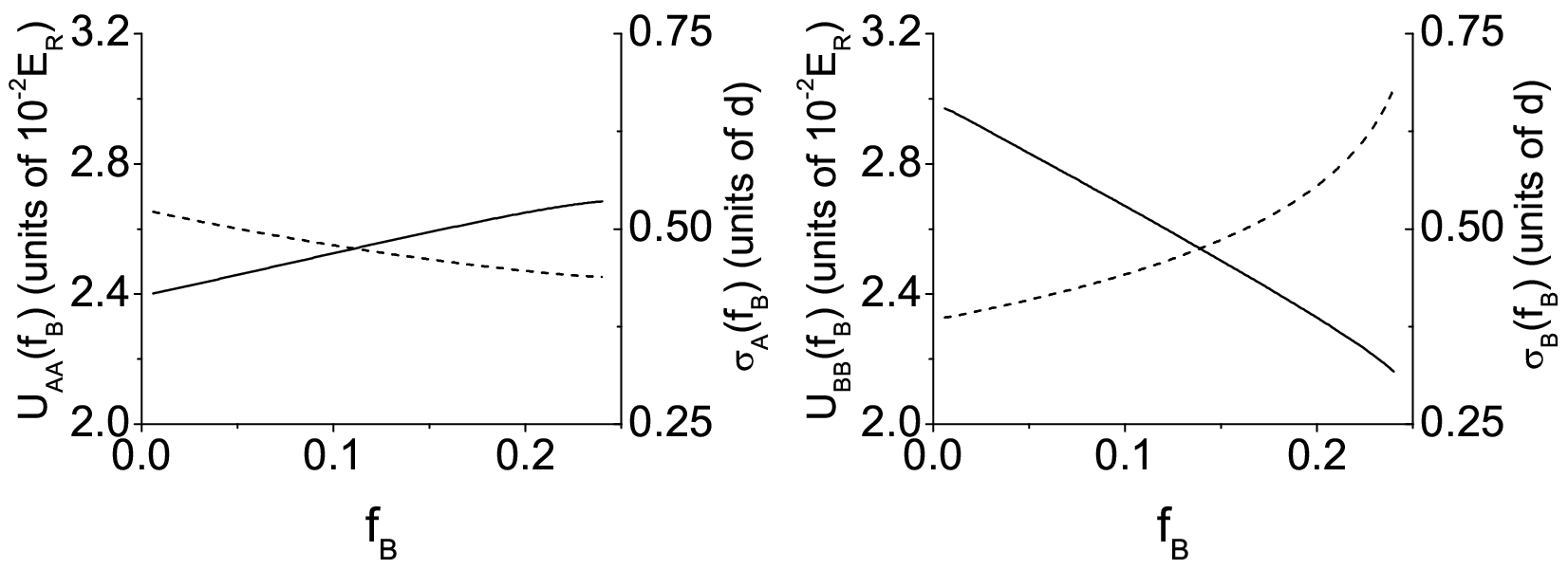}
\caption{The on-site interaction energies, $U_{ii}$ (solid line) and the widths of single-particle wavefunctions, 
$\sigma_i$ (dashed line) for component A 
(left) and for component B (right), showing opposite changes as
the impurity component B populates increasingly. Here $s_{max,A}$=10,
$N_{tot} = 5 \times 10^{3}$. \label{ppl_uwa1}}
\end{figure}

In Fig. \ref{ppl_uwa1}, we show on-site interaction energies for component A
and B as a function of the fraction of component B ($f_{B}$) using the same
parameters as in the TWA simulations, $s_{max,A} =10$ and $N_{tot}=5 \times
10^{3}$.  Higher on-site energies, $U_{ii}$ correspond to greater localization,
(smaller $\sigma_{i}$, where $\sigma_{i}$ is the width of the wavefunction
in the $i$th well) and reduced nonlocal coherence, $C_{i,01}$.

We now analyze phase coherence of component B. We begin with a bosonic mixture 
with a low population of component B, $f_{B} \simeq 0$, which can 
be approximated by the foreground component A with a B impurity. The average 
strength of interspecies interaction per B field ($\sim g_{AB}f_AN$) over its 
spatial variation, is greater when $f_B \simeq 0$ than when $f_B \simeq 1$.
The interaction strength varies over space because of the component A's
modulational variance. This periodic mean field acts similarly to an optical 
lattice for component B, decreasing its phase coherence. On the other hand, 
the mixture with a high population of component B, $f_{B} \simeq 1$ ($q < 1$), 
has weakened phase decoherence, which we can qualitatively interpret by the 
decreased strength of the mean-field 
lattice formed by the component A.

We now analyze phase coherence of component A. For a bosonic mixture with a 
low population of component B, $f_{B} \simeq 0$, the optical lattice alone 
does not substantially induce loss of phase coherence of component A. 
As $f_B$ increases, however, component A loses more phase coherence. The 
larger phase decoherence of component A as $f_A \rightarrow 0$ can be 
understood by the broadening of component B distribution enhanced by the 
narrowing of the A distribution.  Due to the repulsive nature of interspecies 
interaction, the minimum energy is found in the balance between reducing the 
spatial overlap of the two-components' amplitudes and weakening the 
intraspecies interaction energies of each component. 

\subsection{A single lattice (A) with varying heights}\label{both_frag}

In this section, we show how the phase coherence changes as a function of 
time, depending on the final lattice height for component A, in order to see 
the effect of mean-field lattice height. As in Sec. \ref{sec:phcoh_frac},
$V_{o,B}(z,t)=0$, but now the atom numbers are fixed at $N_A=4.0 \times 10^3, 
N_B=1.0 \times 10^3$. As before, the optical lattice for component A linearly
increases up to the indicated value of $s_{max,A}$: the ramping-up
time is $\omega_{R} \tau_{RU} = 100$ in this case.

\begin{figure}
\includegraphics[scale=0.3]{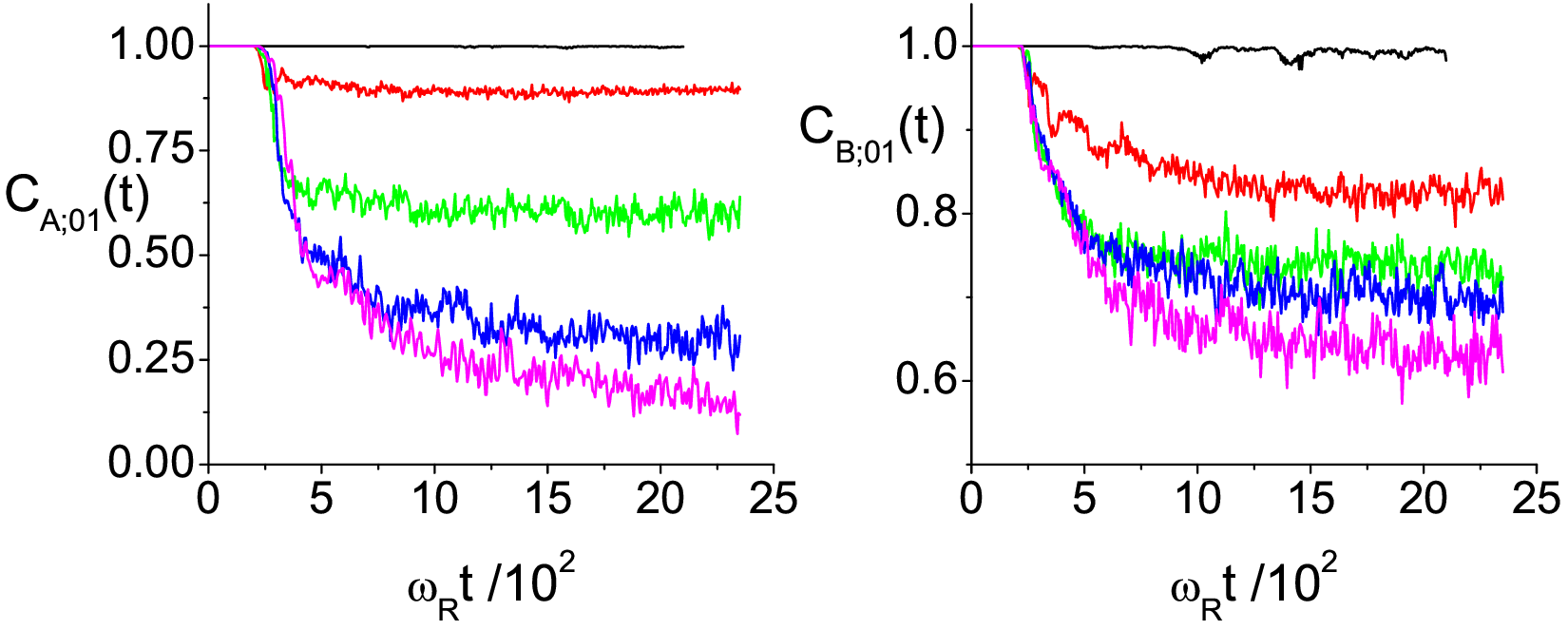}
\caption{(Color online) The effect of final optical lattice heights on phase 
coherence of component A (on left) and component B (on right). The final 
lattice height ($s_{max,A}$) for each curve is following (top to bottom): 
the black line ($3$), the red line ($6.5$), the green line ($10$), the blue 
line ($13.5$), and the pink line ($17$).\label{ppl_av1}}
\end{figure}

\begin{figure}
\includegraphics[scale=0.30]{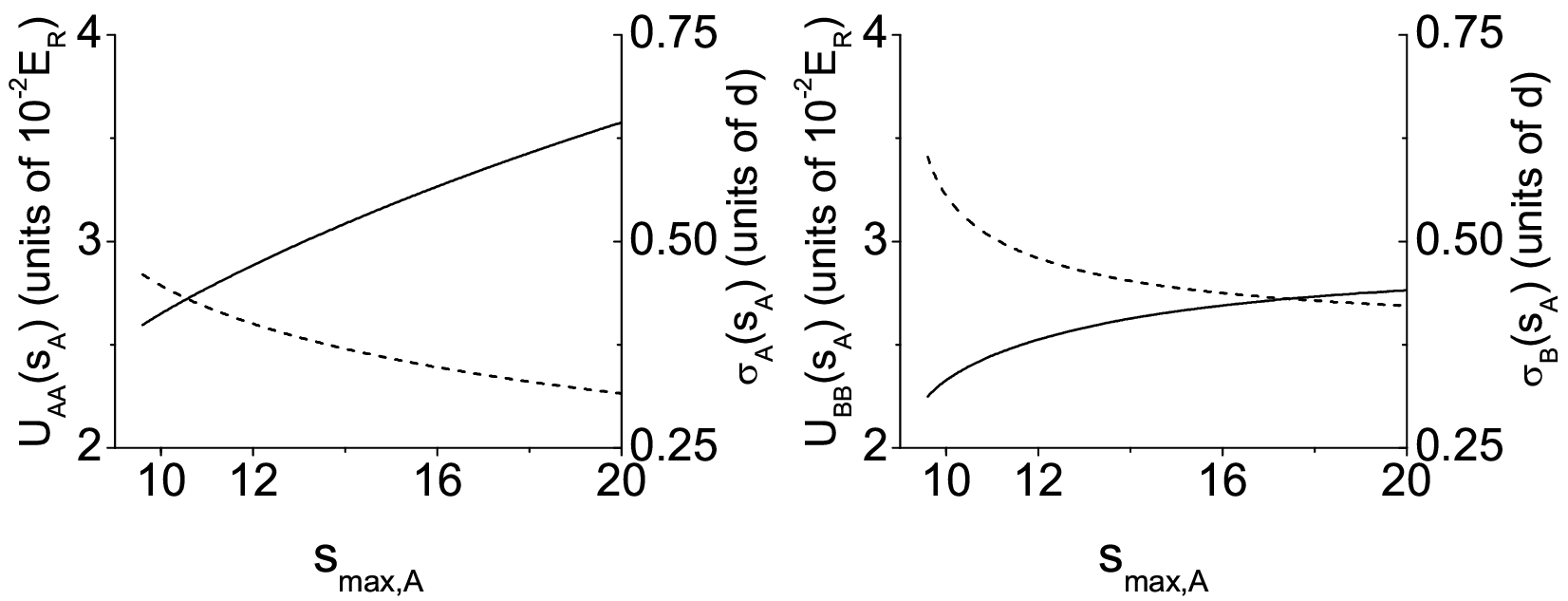}
\caption{The on-site interaction energies, $U_{ii}$ (solid line) and the widths of single-particle wavefunctions, 
$\sigma_i$ (dashed line) for component A ($U_{AA}$ on 
left) and for component B ($U_{BB}$ on right) for 
$N_{tot} = 5 \times 10^{3}$. Both components become more fragmented as the 
state-dependent optical lattice for only component A has been amplified 
more.\label{ppl_uwav1}}
\end{figure}

The changes in phase coherence, $C_{i;01}$, are shown in Fig. \ref{ppl_av1} 
for component A and B, and in Fig. \ref{ppl_uwav1}, the on-site 
interaction energies are displayed for both components.
As the lattice becomes deeper, the phase coherence of 
component A decreases as expected, since it is fragmented by the external 
lattice height increase even without consideration of interspecies effect. 
Decoherence of component B is enhanced as well because of the growth of
the mean-field lattice from component A.  

In light of interaction energies, an increase in the energy implies
a smaller $\sigma$, similarly as seen in Fig. \ref{ppl_uwa1}, hence tighter 
localization within the effective well.
Thus Fig. \ref{ppl_uwav1} indicates, as expected, that the degree of 
localization is higher for deeper lattice heights. Comparing results between 
the two components in Fig. \ref{ppl_uwav1}, the localization of component A is 
evidently stronger than component B for $s_{max,A} > 10$, which explains the 
greater phase decoherence in component A than in component B in 
Fig. \ref{ppl_av1}. 

As is evident from comparing Figs. \ref{ppl_uwav1} and \ref{ppl_uwa1}, when
the A lattice height rises, the exchange of spatial occupation between the two 
components that has been observed in Sec. \ref{sec:phcoh_frac} does not occur.
Component B's localization is strengthened as well as component 
A's. 
The loss in first-order spatial correlation between wells can be induced by 
increasing the height of barriers \cite{Spekkens99}, which for B atoms are 
provided by atomic mean-field lattices in this case.

\subsection{Two peak-mismatched lattices (A, B)}\label{sec:nonmono}

Up to this point, component B has not been subjected directly to an optical 
lattice, but is localized simply by interaction with the mean field resulting 
from component A, and the interspecies interaction.  Additional insight
into the localization process can come from applying to component B an 
optical lattice 
so as to strengthen the localization effect on B atoms on top of the 
former mean field. 
In other words, this section examines the dependence of phase decoherence of 
one component on the fragmentation of the other component in the presence of 
two peak-mismatch optical lattices as in \cite{shi08,gubeskys07}. 
Intuitively, the addition of an optical lattice would additionally increase 
phase decoherence without limit. We will show in this section that it is not 
always the case.

To the BEC mixture with the asymmetric population ratio 
($N_A=1.0 \times 10^3, N_B=4.0 \times 10^3$), we gradually apply two 
state-dependent optical lattices:
\begin{eqnarray}
V_{o,A}(z,t) &=& s_A(t) E_R \cos^2(kz), \nonumber\\
V_{o,B}(z,t) &=& s_B(t) E_R \sin^2(kz),
\end{eqnarray}
where $s_i(t) = s_{max,i} t/\tau_{RU}$. The final lattice height for the 
component A is $s_{max,A}=10$ and the final lattice height for the component 
B is a variable parameter in different simulations ranging from 
$s_{max,B}=3$ to $s_{max,B}=17$ and all other conditions are the same as in 
the previous section.

\begin{figure}
\includegraphics[scale=0.3]{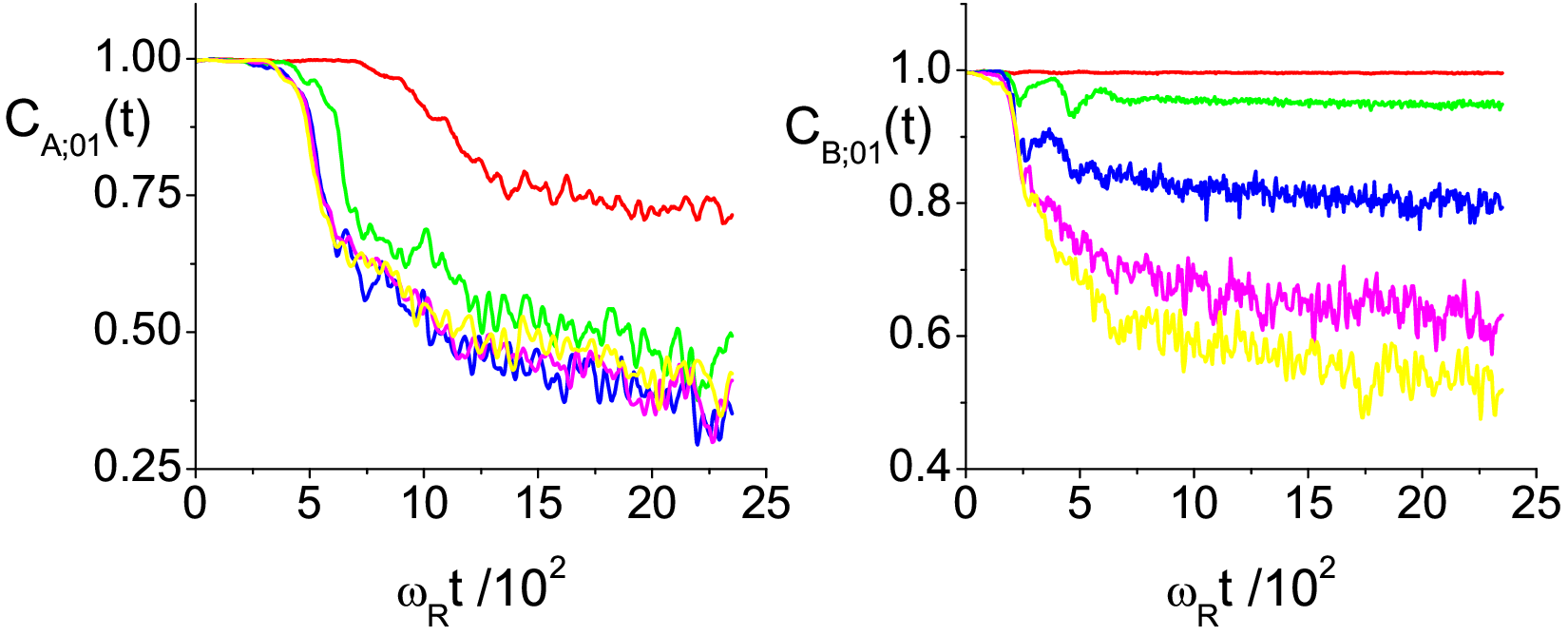}
\caption{(Color online) The effect of localization induced by B on phase 
coherence on component A (on left) and component B (on right). The line 
colors and the final lattice heights for B ($s_{max,B}$) are following 
(explained below in parentheses for the left figure; top to bottom on the 
right): the red line, $3$ (top); the green line, $6.5$ (second to top); the 
blue line, $10$ (one of the two bottom curves); the pink line, $13.5$ (the 
other one of the two bottom curves); the yellow line, $17$ (between the 
bottom and the second to top).\label{ppl_lv1}}
\end{figure}

\begin{figure}
\includegraphics[scale=0.30]{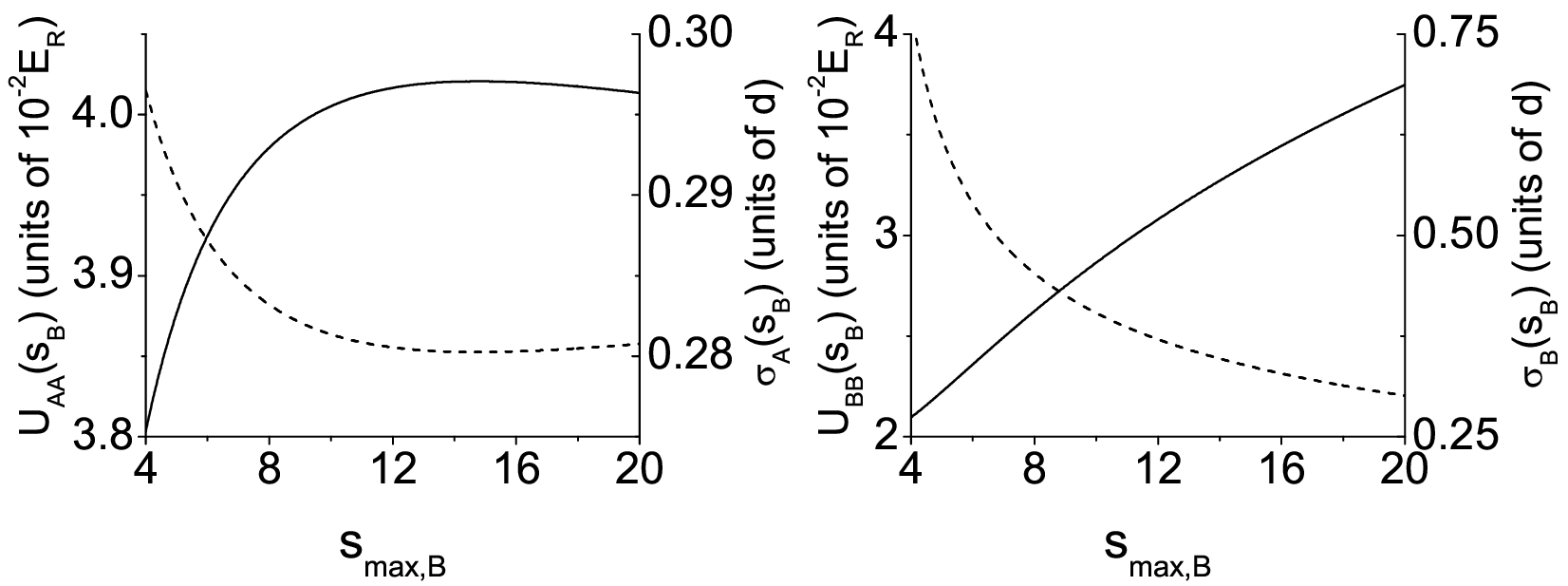}
\caption{The on-site interaction energies, $U_{ii}$ (solid line) and the widths of single-particle wavefunctions, 
$\sigma_i$ (dashed line) for component A ($U_{AA}$ on left) and 
for component B ($U_{BB}$ on right) as a function of the lattice 
height for B ($s_{max,B}$). \label{ppl_uwlv1}}
\end{figure}

In Fig. \ref{ppl_lv1}, we show the change of phase coherence while 
varying the lattice height for component B. The drop in phase coherence of 
component B reflects the expectation that higher optical lattices induce 
more coherence loss between two neighboring sites. And also as expected, 
a higher lattice for component B leads to more localization for component B.

A new phenomenon here is the dependence of phase coherence of component A 
on optical lattice heights for the component B. As can be seen 
from Fig. \ref{ppl_lv1} (left side), the phase coherence of component A at the maximum
time shown is diminished as $s_{max,B}$ increases from 3 to 10, but then
rises again for $s_{max,B}$ increases beyond 10.  This non-monotonic 
dependence on the other component's lattice height can also be seen 
in the on-site energy plotted in the left panel of Fig. \ref{ppl_uwlv1}, which
shows a maximum of $U_{AA}$ at $s_{max,B} \sim 15$, and then a small decrease. 
In addition to the expected fragmentation due to the optical lattice applied
directly to component A, the deepening mean-field potential from component B 
acting on component A further reduces the tunneling of component A between 
adjacent sites until the tunneling of component A suppressed by increased 
mean-field lattice height of component B becomes gradually freed by decreased 
wall width of the mean-field lattice ($\sigma_B$).

\section{Conclusion}\label{sec:conclusion}

In summary, we have investigated nonlocal intraspecies phase decoherence of 
two-component BECs under the gradual loading of state-dependent optical 
lattices. We have used the TWA and the Gaussian variational wavefunctions 
for Wannier orbitals to model the dynamical behavior and fragmentation of 
two species of atoms, in 
particular to calculate the reduction of phase coherence between wells.

First, when a single optical lattice acts on component A 
($V_{o,A} \propto \cos^2(kz), V_{o,B}=0$), the B atoms are fragmented by the 
effective barriers produced by the interspecies repulsive interactions with 
A atoms.
Thus, both species are fragmented into localized wells whose sites differ by 
a half-period between the two components. With varying fractions of a 
mixture, 
we have observed that when the fraction of B atoms increases, then in the 
long-time limit, the fragmentation of A atoms increases, but the 
fragmentation of B 
atoms decreases, consistent with experimental observations in
\cite{catani08,gadway10}. The increasing fragmentation of A atoms is associated
with higher effective barriers produced by accumulation of B atoms spatially
in phase with the A's optical lattice. 
The decreasing fragmentation of B atoms is due to lower effective barriers 
produced by the reduced A mean-field lattice.
With varying heights of the lattice, we have seen increasing fragmentation 
of both A and B atoms with higher lattice heights for A. We confirm that 
deep lattices for only a single component can play a key 
role for both components in reducing phase coherence.

Finally, when optical lattices are applied to both components with the 
condition that their well sites differ by a half-period 
($V_{o,A} \propto \cos^2(kz), V_{o,B} \propto \sin^2(kz)$), then as the 
height of one of the lattices increases, phase decoherence of A atoms is 
limited and non-monotonic fragmentation occurs;  
the A atom on-site interaction energies reach a maximum, and then decrease 
while the other atoms (B) become more localized. This shows in a dramatic way 
how the effect of fragmentation (or equivalently, phase decoherence) of one 
species due to other component's mean-field, can actually saturate.

All the calculations pertain to a situation in which the two species are 
different hyperfine states of the same atomic species, with equal masses and 
equal inter- and intra-species scattering lengths. Since this work is 
exploratory in nature, we have not attempted an extensive survey of 
parameter space by varying atom numbers, masses, and scattering lengths. We 
suggest that effects similar to what we obtain could occur if the two species 
were actually different types of atoms (Rb and K, for example), such as has 
been obtained in recent experiments. 

\begin{acknowledgments}
This work was supported by the US NSF under grants PHY0652459 and PHY0968905, 
and by the research fund of Hanyang University (HY-2014-N). We also thank 
Prof. Dominik Schneble for a careful and critical reading of this paper, 
and B. Gadway for helpful comments. We are also indebted to Prof. Hong Ling 
of Rowan University for clarifying theories and making detailed and 
extensive comments. 
\end{acknowledgments}

\appendix

\section{Initial states and TWA for two-component BECs}\label{sec:TWA}

First, we generate a set of classical stochastic fields for the
initial state sampled from the corresponding Wigner distribution 
function \cite{gardiner04, isella06}
and obtain the dynamics of the system by averaging the statistical ensemble
over individual trajectories in phase space. The expectation values of
symmetrically-ordered operators are calculated from the weighted average of the
corresponding classical fields ($\psi_W$) with the Wigner distribution
function, $W(\psi_W,\psi_W^*)$ without extra modification terms.

The classical stochastic fields, $\alpha_{A0}, \alpha_{B0}, \alpha_{\mu}$ 
are obtained by randomly generating c-numbers that follow a given
distribution, corresponding to quantum
operators, $\hat{\alpha}_{A0}, \hat{\alpha}_{B0}$, and $\hat{\alpha}_{\mu}$.
Since the
condensate mode operators and quasiparticle mode operators commute and the
component A and B condensate operators also commute, we independently sample
the c-numbers. For the condensate mode, the initial two-component superfluid
state is approximated as a coexisting mixture of
independent Glauber coherent states, where each coherent state preserves its
own phase coherence. The Wigner function for the initial state is
\begin{eqnarray}
W(\vec{\alpha},\vec{\alpha}^*) &=& 
W_{A0}(\alpha_{A0},\alpha_{A0}^*)W_{B0}(\alpha_{B0},\alpha_{B0}^*) \nonumber\\
&& W_{BG}(\alpha,\alpha^*),
\end{eqnarray}
where $\vec{\alpha} = (\alpha_{A0}, \alpha_{B0}, \alpha_\mu)^T$ and $W_{A0},
W_{B0}, W_{BG}$ are the Wigner distribution for the component A, the component
B, and the Bogoliubov modes. The condensate mode Wigner distributions are
given by ($i$ = A, B)
\begin{eqnarray}
W_{i0}(\alpha_{i0},\alpha_{i0}^*)&=&\frac{2}{\pi}\exp{\big[-2|\alpha_{i0}
-\sqrt{N_{i0}}|^2\big]},
\end{eqnarray}
where the ensemble averages are $\langle\alpha_{i0}\rangle_W=\sqrt{N_{i0}}$
and $\langle\alpha_{i0}^*\alpha_{i0}\rangle_W=N_{i0}+\frac{1}{2}$, and W
denotes the ensemble average in the Wigner distribution.
The distribution function of coherent states has a Gaussian
profile in the complex phase space with a variance of $1/2$. For a
large number of atoms ($N \gg 1$), quantum fluctuations around the mean
classical field are relatively small, since $\Delta N/\langle N \rangle =
1/\sqrt{\langle N \rangle}$. Thus we can think of the initial state as a
classical field with a small fluctuation in phase space.

While the condensate modes have nonzero expectation values for the atom 
number of components, $\alpha_{i0}$, the noncondensate modes, $\alpha_{\mu}$
have zero expected populations, for which the Wigner distribution is the
product of uncorrelated Wigner functions for each mode \cite{blakie08}:
\begin{eqnarray}
W_{BG}({\alpha,\alpha^*})&=&\prod_{\mu}W_{\mu}(\alpha_{\mu},\alpha_{\mu}^*), 
\nonumber\\
W_{\mu}(\alpha_{\mu},\alpha_{\mu}^*)&=&
\frac{2}{\pi}\tanh{\Big(\frac{\epsilon_{\mu}}{k_BT}\Big)} \nonumber\\
&& \exp{\bigg[-2|\alpha_{\mu}|^2\tanh{\Big(\frac{\epsilon_{\mu}}{k_BT}}\Big)
\bigg]},
\end{eqnarray}
where $W_{BG}({\alpha,\alpha^*}), W_{\mu}(\alpha_{\mu},\alpha_{\mu}^*)$ is
the Wigner function for the total Bogoliubov modes and for each quasiparticle 
mode, respectively. The ensemble averages of quasiparticle modes satisfy the
condition that they have zero mean values and Gaussian variances,
which broaden as the temperature increases.
\begin{eqnarray}
\langle\alpha_{\mu}\rangle_W=\langle\alpha_{\mu}^*\rangle_W=0, \\
\langle\alpha_{\mu}^*\alpha_{\nu}\rangle_W=\delta_{\mu\nu}[n_{\nu}+\frac{1}{2}].
\end{eqnarray}

Then, we construct a TWA method for the dynamics of two-component BECs under the
nonequilibrium ramp-up of state-dependent optical lattices. We begin by 
projecting the above two-component state onto its phase space within the
Wigner representation. We thereby obtain a quasi-probability
distribution function over the phase space, which is formulated to be
analogous to the density matrix in quantum mechanics \cite{baker58,steel98}. 
Even though a positive-P representation is sometimes used for simulations of a
quantum system, it is subject to instabilities, for example in highly
populated modes \cite{steel98}. Instead, we approximate the dynamics of 
one-dimensional trapped multimode BECs by considering the Fokker-Planck 
equation in the truncated Wigner representation \cite{steel98}.

The Fokker-Planck equation for the Wigner quasiprobability distribution yields
the time evolution equation:
\begin{eqnarray}
\frac{\partial W(\vec{\psi}, \vec{\psi}^*)}{\partial t}=
\int dz \frac{i}{\hbar}\sum_{i,j=A,B}
 \bigg[\frac{\delta}{\delta\psi_i(z)}\Big\{L_i
\hspace*{1cm} \phantom{aaaa} \nonumber \\
+ g_{ij}(|\psi_j(z)|^2-d_{ij})\Big\}\psi_i(z) \phantom{aa} \nonumber\\
 -\frac{g_{ij}}{4}\frac{\delta}{\delta\psi_i(z)}
\frac{\delta}{\delta\psi_j(z)}\frac{\delta}{\delta\psi_j^*(z)} 
\psi_i(z)\bigg]W(\vec{\psi},\vec{\psi}^*) + h.c.,\label{eq:FP} \phantom{aa}
\end{eqnarray}
where $\vec{\psi} = (\psi_A, \psi_B)^T$, and 
$d_{ij} = 1 \phantom{a} (\mathrm{or} \phantom{a} 1/2)$ 
if $i=j \phantom{a} (\mathrm{or} \phantom{a} i \ne j)$. 

The exact Fokker-Plank equation with the presence of the third-order term
within the Wigner representation is difficult to solve both analytically and
numerically in stochastic simulations \cite{sinatra02}. Therefore, the
truncation in the TWA neglects the third-order derivative
terms in Eq. (\ref{eq:FP}), which are smaller than the Gross-Pitaevskii
first-order term in the total number, $N$. The second-order diffusion process 
term of usual stochastic processes, which can have a prominent role in 
enhancing fluctuations, is absent in the TWA. Also, in the TWA, the
expectation values of those classical fields,
$\prod_{i,j}\psi_i^*(z_i)\psi_j(z_j)$ in the Wigner representation,
correspond to the expectation values of quantum operators that are
symmetrically ordered. The Wigner quasiprobability distribution
function, $W(\vec{\psi},\vec{\psi}^*)$, is a classical projection function
corresponding to the density operator for the field operators in quantum
mechanics:
\begin{eqnarray}
\langle\prod_{i,j}\hat{\psi}_i^\dagger(z_i)\hat{\psi}_j(z_j)\rangle_W=
\nonumber\\
\int d^2\vec{\psi} W(\vec{\psi},\vec{\psi}^*)\prod_{i,j}\psi_i^*(z_i)
\psi_j(z_j).
\end{eqnarray}

\section{The Gaussian ansatz for Wannier functions}\label{sec:Gaussian_Wannier}

We expand the
wavefunctions in an array of Wannier-like orbitals, $w_i(z)$,
\begin{eqnarray}
\hat{\psi}_i(z) = \sum_{l} \hat{a}_{il} w_i(z-R_{il}),
\end{eqnarray}
where the single particle wavefunction, $w_i(z-R_{il})$ is centered at
$R_{Al} = (2l\pm 1)d/2$, $R_{Bl}=2ld/2$ for each component, and the operators
$\hat{a}_{il}$ satisfy the bosonic commutation relation,
$[\hat{a}_{il},\hat{a}_{jl'}^{\dagger}] = \delta_{ij}\delta_{ll'}$. The
variationally minimum solution of orbital wavefunctions implicitly depends 
on the occupation per site. Putting this set of orbitals into the Hamiltonian
in Eq. (\ref{eq:H_sys}), we obtain
\begin{eqnarray}
H &=& - \sum_{i;ll'} J_{i;ll'} (\hat{a}_{il}^{\dagger}\hat{a}_{il'}
+\hat{a}_{il'}^{\dagger}\hat{a}_{il}) + \sum_{i;l}
U_{ii}\hat{a}_{il}^{\dagger}\hat{a}_{il}^{\dagger}\hat{a}_{il}\hat{a}_{il}
\nonumber\\
&& + \sum_{l}U_{AB}\hat{a}_{Al}^{\dagger}\hat{a}_{Bl}^{\dagger}
\hat{a}_{Al}\hat{a}_{Bl} + \sum_{i;l}\epsilon_{il}\hat{a}_{il}^{\dagger}
\hat{a}_{il},
\end{eqnarray}
where
\begin{eqnarray}
J_{i;ll'} &=& -\int dz w_i(z-R_{il}) \nonumber\\
&& \bigg[-\frac{\hbar^2}{2m_i}\bigtriangledown^2 + V_{h,i}(z) + V_{o,i}(z,t)
\bigg] w_i(z-R_{il'}),\nonumber\\
U_{ij} &=& g_{ij}\int dz w_i^2(z-R_{il})w_j^2(z-R_{jl}).
\end{eqnarray}

In the tight-binding limit, the Wannier functions can be written as Gaussian functions \cite{slater52}. When the
tight-binding limits $(V_{o,A}(z,t)|_{max}-V_{o,A}(z,t)|_{min}) \gg E_R$ and
$(I_{al,A}(z)|_{max}-I_{al,A}(z)|_{min}) \gg E_R$ are achieved, the high
vibrational modes for each component are not occupied especially at the
initial temperature $T=0$ so that the profile of each component can be well
described by the ground state, the Gaussian wavefunction.
Starting from the initial trial state of infinite 1D BECs in the
periodic state-dependent optical lattice, we employ the Gaussian variational
ansatz for a single-particle orbital placed on each site,
$w_i(z-R_{il}) = (1/\pi\sigma_{i}^2)^{1/4}\exp(-(z-R_{il})^2/2 \sigma_{i}^2)$,
with the density of atoms per site equal to the average density of the
center site calculated from the GPE ($n_{il} = n_{i0}^{(GPE)}$). Here, the 
widths of Gaussian wavefunctions are variational parameters, as in
\cite{perez-garcia97,salasnich02,vignolo03,Schaff10}.

Within the Gaussian approximation, we obtain the minimized Gross-Pitaevskii
energy functional, where the interaction energies and the tunneling amplitudes
can be calculated from variational parameters. The local single-band
Gaussian state is known to be accurate for the calculation of on-site
interaction energies even for shallow lattices ($\sim 3E_R$) with the overlap
between the true Wannier function and the Gaussian wavefunction nearly equal
to 1.0 \cite{bloch08}. Since the Gaussian ansatz can be quite imprecise for the
calculation of tunneling amplitudes because of the tail of Gaussian
functions \cite{bloch08,kramer03}, we concentrate on calculating
on-site interaction energies.

\section{Numerical method}\label{sec:num_meth}

In this section, we explain the numerical methods implemented in this work. The condensation of cigar-shaped 1D atomic clouds is achieved in an
anisotropic harmonic trap with the tight confinement along the transverse
direction ($\omega_z=2\pi \times 130Hz, \omega_{\rho}=2\pi \times
2.71kHz$) and the aspect ratio is 21. In this work, the number of
atoms ranges from $0.5 \times 10^3$ to $5 \times 10^3$. The optical lattices
are generated by red-detuned off-resonant lasers with a wavelength
$\lambda = 785$ nm, so that the period is $d=\lambda/2 = 392.5$ nm, and the
recoil frequency is $\omega_{R} = (\hbar/2m) (2
\pi/\lambda)^{2} = 2 \pi \times 3.73$ kHz.

With $N_{tot} = 5 \times 10^3$, and the trap and lattice properties given above,
we obtain as many as 75 atoms in the central well of the quasi-1D lattice. 
Experimentally, working also with $^{87}$Rb, Campbell et al. \cite{Campbell06} 
were able to put at most 5 atoms per site in their 3D optical lattice.  
Because of the tighter transverse confinement in \cite{Campbell06}, 
75 atoms per well in our simulations would actually correspond to about 
8 atoms per well in \cite{Campbell06} for the same density at the peak. In 
actual experiments, the total number of atoms would need to be reduced
over the value used here. This would lead to a reduction of the 
demonstrated coherence loss effects.

The one-dimensionality of cigar-shaped BECs in the harmonic trap at $T=0$ is
achieved when $l_z > \xi > l_{\rho}$ or $\mu_{3D} <
\hbar\omega_{\rho}$ \cite{gorlitz01}, where $l_z=(\hbar/m\omega_z)^{1/2}$
and $l_{\rho}=(\hbar/m\omega_{\rho})^{1/2}$ are the longitudinal and the
transverse zero-point oscillation length respectively, 
$\xi = (1/4\pi n_{3D} a_s)^{1/2}$ is the healing length, and $\mu_{3D} =
\hbar^2/2m(15Na_s/l_z^2l_{\rho}^4)$ is the chemical potential corresponding
to the interaction energy. Furthermore at $T>0$, $l_{\rho}$ is required to
be smaller than the de Broglie wavelength $\lambda_T$ ($l_{\rho} <
\lambda_T$), where $\lambda_T = (2\pi\hbar^2/mT)^{1/2}$ \cite{kheruntsyan03}.
For the $^{87}$Rb atoms in a trap with frequencies given above, the ratios,
$l_{\rho}/\xi \lesssim 2$ and $\mu_{3D}/\hbar\omega_{\rho} \lesssim 3$.

Each BEC component lies in the Thomas-Fermi regime between the full 3D
dynamics and the true 1D dynamics with transverse excitations almost frozen
out. Even though the BEC is in the crossover between 3D and 1D, the
low-energy excitation modes in 3D are effectively 1D provided that the
temperature is sufficiently below the energy of the transverse
oscillator ($T < \hbar\omega_{\rho}$) \cite{Stringari98} which is the case 
here ($T=0$). The Thomas-Fermi radius ranges from $5$ to $12 l_z$.

The dimensionless coupling strength of interaction energies in this work
is $\gamma = mg_{1D}/\hbar^2n_{1D} \lesssim 2 \times 10^{-3}$ and the
reduced temperature is $\tau = 2mk_BT/\hbar^2n_{1D}^2 = 0$ \cite{gangardt03}.
Therefore, the 1D Bose gas can be effectively described by the
Gross-Pitaevskii equation in the regime ($\tau^2 \lesssim \gamma \lesssim 1$)
far from the Tonks-Girardeau regime ($\gamma \gtrsim 1$). The nonlinearity
$g_{1D}N/\hbar\omega_zl_z$ \cite{isella06} ranges from $120$ to $1200$.

The numerical preparation of initial states requires ground state
wavefunctions,
the Bogoliubov quasiparticle excited modes, and their stochastic distributions
governed by the Wigner functions. We find the ground state wavefunctions by
numerically integrating the GPE in imaginary time with a time step of
$\omega_R\delta t=0.005$ with 3072 spatial grid
points along the axial direction. We utilize the second-order split-operator
method to integrate the time evolution of wavefunctions, in nonlinear as
well as linear regimes.  Using the ground state solutions of the GPE, we obtain
quasiparticle wavefunctions, $u_{A\mu}(z),u_{B\mu}(z),v_{A\mu}(z),v_{B\mu}(z)$
for energy, $\epsilon_{\mu}$ by the diagonalization of the
Bogoliubov-de Gennes equation [Eq. \ref{eq:BdG}].

We calculate the ensemble average of stochastic fields along the trajectory
and find their coherence. Stochastic quantum fluctuations are appended to the
initial mean-field state for generation of the ensemble of
Wigner-distributed initial states, in which step we perform the Gaussian
random variable generation of order parameters ($\alpha_{i0},\alpha_{\mu}$). 
For the condensate mode, the mean of $\alpha_{i0}$ is
$\sqrt{N_{i0}}$ and its width of deviation is $\sqrt{1/2}$, whereas for the
Bogoliubov quasiparticle mode, the mean of $\alpha_{\mu}$ is zero and the width
is $\sqrt{1/2}$ for $T=0$. A single sample of stochastic fields,
$\psi_W(x)$ is obtained by configuring the wavefunction profiles with the
generated stochastic order parameters.

The condition for numerical validity of the TWA method in the Bogoliubov
theory is that the condensate mode must be highly populated compared to the
noncondensate mode so that the quantum fluctuation is small, being dominated 
by the condensate field. In other words, the TWA in the mean-field theory is 
valid with a relatively small number of excited Bogoliubov quasi-particles 
compared to the number of condensate particles in the system, $N \gg M/2$, 
where N is the total number of atoms, M is the number of Bogoliubov 
quasi-particles. This is a regime different from other exact numerical 
methods, for example, the Time Evolving
Block Decimation (TEBD) method or the Density Matrix Renormalization
Group (DMRG) with the Bose-Hubbard Model, in which cases each site is limited
to a low filling factor since the Hilbert space increases exponentially with
the number of atoms and the number of sites.

We perform the simulation with an ensemble of states consisting of 500
samples for the TWA distribution function to achieve sufficient convergence.
The time evolution of ensembles has the typical time step given by
$\omega_R \delta t=0.005$, i.e. $\delta t$ = 0.2 $\mu$s.

\bibliography{tlib3}

\end{document}